\documentclass[12pt]{iopart}
\usepackage{graphicx}
\usepackage{float}
\usepackage[geometry]{ifsym}
\usepackage{color}
\usepackage{hhline}
\usepackage{array}
\usepackage{multirow}
\graphicspath{{./Figures/}}

\begin{document}
\title{Control of the detached flow behind a backward-facing step by visual feedback}
\author{N. Gautier and J.-L. Aider}
\address{PMMH, 10, rue Vauquelin 75006 Paris, France}

\begin{abstract}
The separated flow downstream a backward-facing step is controlled using visual information for feedback. This is done when looking at the flow from two vantage points. Flow velocity fields are computed in real-time and used to yield inputs to a control loop. This approach to flow control is shown to be able to control the detached flow in the same way as has been done before by using the area of the recirculation region downstream the step as input  for a gradient descent optimization scheme (\cite{King2007}). Visual feedback using real-time computations of 2D velocity fields also allows for novel inputs in the feedback scheme. As a proof of concept, the spatially averaged value of the swirling strength $\lambda_{ci}$ is used successfully as input for an automatically tuned PID controller.

\end{abstract}


\section{Introduction}
Flow separation is a subject of scientific and industrial importance. Separated flows can strongly influence the performances for many industrial devices, such as diffusers, airfoils, air conditioning plants, moving vehicles, and many other industrial systems (\cite{CHC,Hucho2005,Weisenstein2000}). Usually the focus is on minimizing the size of the recirculation region, also called recirculation bubble, to improve drag, reduce lift, reduce vibrations or lower aeroacoustic noise.

To manipulate a detached flow many kinds of actuations have been considered. A list of actuations can be found in \cite{Fernholz1990}. In the case of open-loop flow control,  passive actuations often improve characteristics of the flow only for given operating conditions. Active actuators have also been used in open-loop  experiments, (see \cite{Chun1996,Mazur2007}), but they are unable to adapt to exogenous parameter changes.  Feedback control strategies, or closed-loop flow control, offer the possibility of adapting the actuation to external perturbations or changes in the experimental conditions, thus improving the robustness of the control. One can cite a few recent examples of closed loop control strategies implemented either numerically or experimentally: \cite{Beaudoin2006,King2007,Pastoor2008}.
\\
 The backward-facing step (BFS) is considered as a benchmark geometry for studying detached flow. Separation is imposed by a sharp edge, thus allowing the separation process to be examined by itself. The main features of the BFS flow are the creation of a recirculation bubble downstream of the step together with a strong shear layer in which Kelvin-Helmholtz instability can trigger the creation of spanwise vortices. The flow downstream a backward-facing step has been extensively studied  both numerically and experimentally, see \cite{Armaly1983,Le1997,JLA2004,JLA2007}.
\\
An essential part of many control strategies is determining one or several control variables. The variable is either directly computable from sensor data, such as local pressure or drag measurement, or obtained by combining sensor data and a model. This model can be simple (\cite{King2007} recover recirculation length via its correlation to pressure fluctuations) or complicated (\cite{Sipp2010} recovers an approximation of the flow state through Kalman filtering). In the case of the backward-facing step, sensors are almost always pressure sensors, and the control variable is usually the recirculation bubble length. While wall based sensors present the advantage of high frequency acquisition they also present a limited view of the flow: many phenomena are difficult to access because buried in noise or simply unobservable. Furthermore they are intrusive.
\\
Using the flow velocity field computed from visual data to control a flow has been suggested and successfully implemented in numerical simulations (\cite{Collewet2011}). Control using visual feedback was implemented as a proof of concept by \cite{Willert2010}. \cite{Roberts2012} was successful in improving the control of the flow behind a flap using real-time instantaneous velocity data. The exponential rise in computing power allows for the computation of large, dense, accurate velocity fields at high frequencies as shown in \cite{Gautier2013OF}.
\\
In this paper we investigate the feasibility of controlling the separated flow behind a backward-facing step using flow velocity fields computed in real-time from visual data. The flow is observed both from the side and from above. Two control variables are used, the recirculation bubble surface and the spatially averaged swirling intensity. A jet is used to act on the flow. The evolution of the control variables with Reynolds  number (based on step height $h$) or the jet velocity is determined. Two basic schemes are implemented to control the flow. To show how visual feedback can be used in the same control schemes as have been previously proposed the recirculation bubble surface  is used as input variable to a gradient descent optimization scheme. To show visual servoing can be used to control the flow in novel ways, the control variable based on swirling strength is  used as input to a PID controller. The aim of this paper is to show the relevance of visual feedback for detached flows. Thus the injection geometry, control variables and control schemes are reduced to their simplest expression.

\section{Experimental setup}

\subsection{Water tunnel}
Experiments were carried out in a hydrodynamic channel in which the flow is driven by gravity.
 The walls are made of Altuglas for easy optical access from any direction. The flow is stabilized by divergent and convergent sections separated by honeycombs. The test section is $80$~cm long with a rectangular cross section $15$~cm wide and $10$~cm high.\\
 
 The mean free stream velocity $U_{\infty}$ ranges between $1.38$ to $22$~cm.s$^{-1}$.  A specific leading-edge profile is used to start the boundary layer smoothly which then grows downstream along the flat plate, before reaching the edge of the BFS. The boundary layer has a shape factor $H \approx 2$. The quality of the main stream can be quantified in terms of flow uniformity and turbulence intensity.  The standard deviation $\sigma$ is computed for the highest free stream velocity featured in our experimental set-up. We obtain $\sigma = 0.059$~cm.s$^{-1}$ which corresponds to turbulence levels of $\frac{\sigma}{U_{\infty}}=0.0023$.

\subsection{Backward-facing step geometry}
\begin{figure}[H]
\centering
\includegraphics[width=0.7\textwidth]{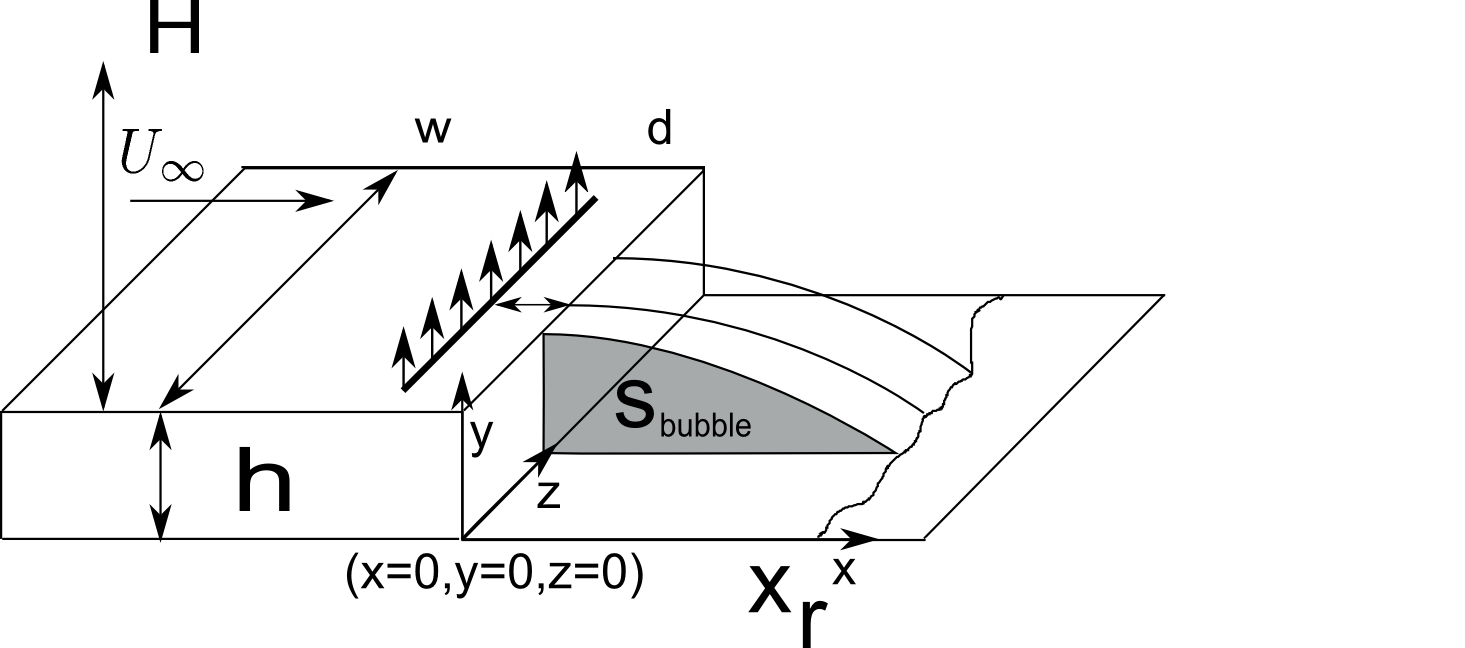}
\caption{Sketch of the BFS geometry and definition of the main parameters.}
\label{fig:dimensions}
\end{figure}

The backward-facing step geometry and the main geometric parameters are shown in figure~\ref{fig:dimensions}. The height of the BFS is $h=1.5$~cm, leading to Reynolds numbers $Re_h = \frac{U_{\infty}h}{\nu}$ ranging between 0 and 3300 ($\nu$ being the kinematic viscosity). Channel height is $H=7$~cm for a channel width $w=15$~cm. Then, one can define the vertical expansion ratio  $A_y = \frac{H}{h+H} = 0.82$ and the spanwise aspect ratio $A_z=\frac{w}{h+H}=1.76$. The distance between the injection slot and step edge is $d=35$~cm.

\begin{figure}[H]
\centering
\includegraphics[width=1.1\textwidth]{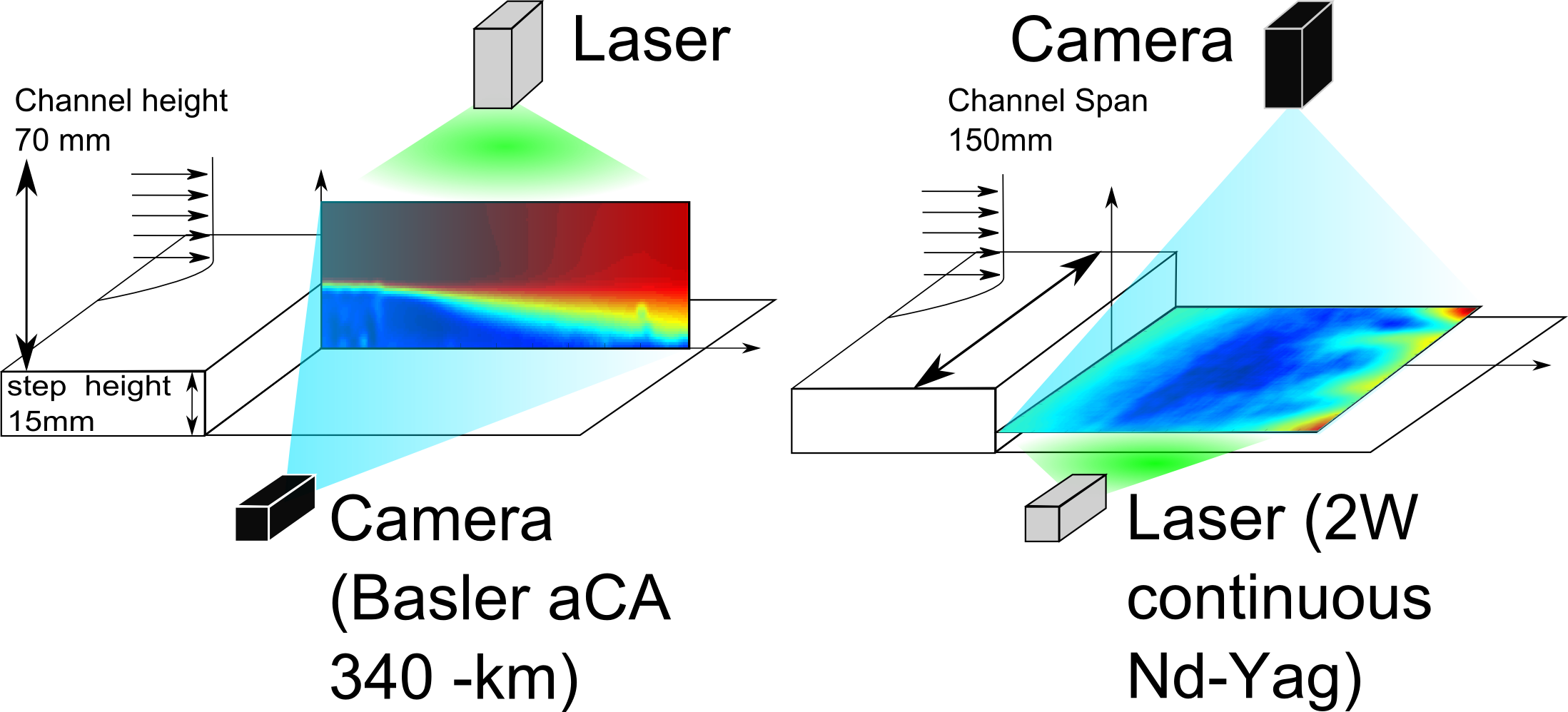}
\caption{Definition of the two measurement configurations used to characterize the separated flow downstream the backward-facing step: measurements in the symmetry plane (SP) and in a $y=0.5$~cm horizontal plane (HP).}
\label{fig:Sketch_side_above}
\end{figure}

\subsection{Real-time 2D2C velocimetry}

The flow is seeded with 20~$\mu m$  neutrally buoyant polyamid seeding particles.  The test section is illuminated by a laser sheet created by a 2W continuous CW laser operating at wavelength $\lambda = 532$~nm and a cylindrical lens.  The pictures of the illuminated particles are recorded using a Basler acA 2000-340km 8bit CMOS camera. The camera is controlled by a camera-link NI PCIe 1433 frame grabber allowing for real-time acquisition and processing. Velocity field computations are ran on the Graphics Processor Unit (GPU) of a Gforce GTX 580 graphics card. 
 
The 2D2C (measurements of two components in a 2D plane) velocimetry measurements are obtained using an optical flow algorithm running entirely on a GPU. It is a local iterative gradient-based cross-correlation optimization algorithm which yields dense velocity fields, i.e. one vector per pixel.  It belongs to the Lucas-Kanade family of optical flow algorithms (\cite{Lucas1984}). Its offline accuracy was extensively studied by \cite{Plyer2011} and its online efficacy at high frequencies was demonstrated in \cite{Gautier2013OF}. 
The spatial resolution however is tied to the window size, like any other window based PIV technique. However the dense output is advantageous since it allows the sampling of the vector field very close to obstacles, yielding good results near walls, as shown in \cite{Plyer2011}.

\begin{figure}
\centering
\includegraphics[width=0.40\textwidth]{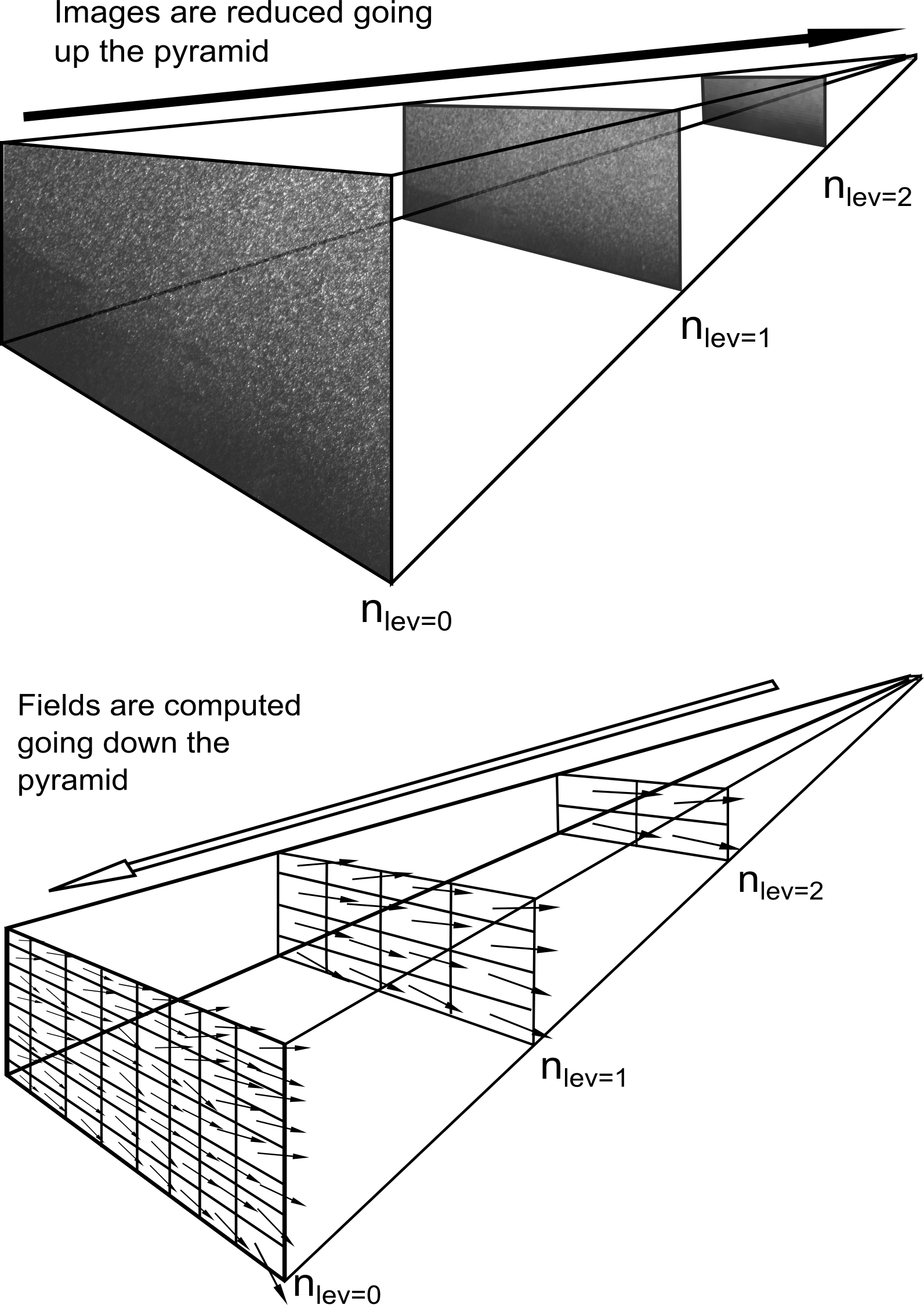}
\caption{Sketch of the computation pyramid.}
\label{fig:pyramid}
\end{figure}

The principle of the featured optical flow algorithm is as follows. The original images are reduced in size by a factor of 4 iteratively until intensity displacement in the reduced image is close to 0. This gives a pyramid of images, described in figure \ref{fig:pyramid}. Displacement is computed in the top image with an initial guess of zero displacement using an iterative Gauss-Newton scheme to minimize a sum of squared difference criterion. This displacement is then used as an initial estimate for the same scheme in the next pair of images in the pyramid. And so on until the base of the pyramid, corresponding to the initial images is reached, thus giving the final displacement. 

It should be noted this pyramidal process allows the algorithm to converge for small and high displacements indiscriminately making it ideally suited to compute velocities in separated flows where regions of low and high velocity fluid often coexist.

Because of its intrinsically parallel nature the algorithm is able to fully utilize the processing power of modern GPUs. It is not uncommon to reach $99\%$ GPU loads. Allowing it to accurately compute flow velocity fields for large images (typically 2 Megapixels) at relatively large frame rates (24 frames per second or greater frequencies with the given hardware). Furthermore it scales exceptionally well with increasing computing power. Although there are differences with classic PIV algorithms, output velocity field resolution is still tied to the size of the interrogation window. Nevertheless, the output field is dense (one vector per pixel) giving better results in the vicinity of edges and obstacles. Furthermore this gives exceptionally smooth fields.

Two configurations are used in the following: the first is the classic vertical symmetry plane (SP) and the second is a horizontal plane (HP), as illustrated in figure \ref{fig:Sketch_side_above}. The position of the horizontal plane, for the HP configuration, was $y = 0.5$~cm. The position was the result of a compromise. The plane was placed as close to the lower wall as possible while still getting good image quality. When the plane is too close to the wall, the light reflected by the particles on the bottom wall makes the measurement difficult. Figure \ref{fig:Sketch_side_above} shows how the flow was observed both from the side and from above. Characteristics of the measurement for both configurations are detailed in table \ref{tab:side_above}.

\begin{table}[H]
\centering
\footnotesize{
\begin{tabular}{c|c|c|}
\cline{2-3}
& Symmetry plane (SP) & Horizontal plane (HP) \\ \cline{1-3}
 \multicolumn{1}{|c| }{ \multirow{1}{*}{ Position} }&  Medium plane & 5mm from bottom \\ \cline{1-3}
\multicolumn{1}{|c| }{ \multirow{1}{*}{ Image resolution} }& 1792x384 & 1920x1024 \\ \cline{1-3}
\multicolumn{1}{|c| }{ \multirow{1}{*}{ mm/pixel} }&  0.107 & 0.07 \\ \cline{1-3}
\multicolumn{1}{|c| }{ \multirow{1}{*}{ Interrogation window size} }& 10x10 & 32x32 \\ \cline{1-3}
\end{tabular}
}
\caption{Characteristics of measurement windows for SP \& HP configurations}
\label{tab:side_above}
\end{table}

\subsection{Actuation \& Feedback loop}
\begin{figure}[H]
\centering
\includegraphics[width=0.6\textwidth]{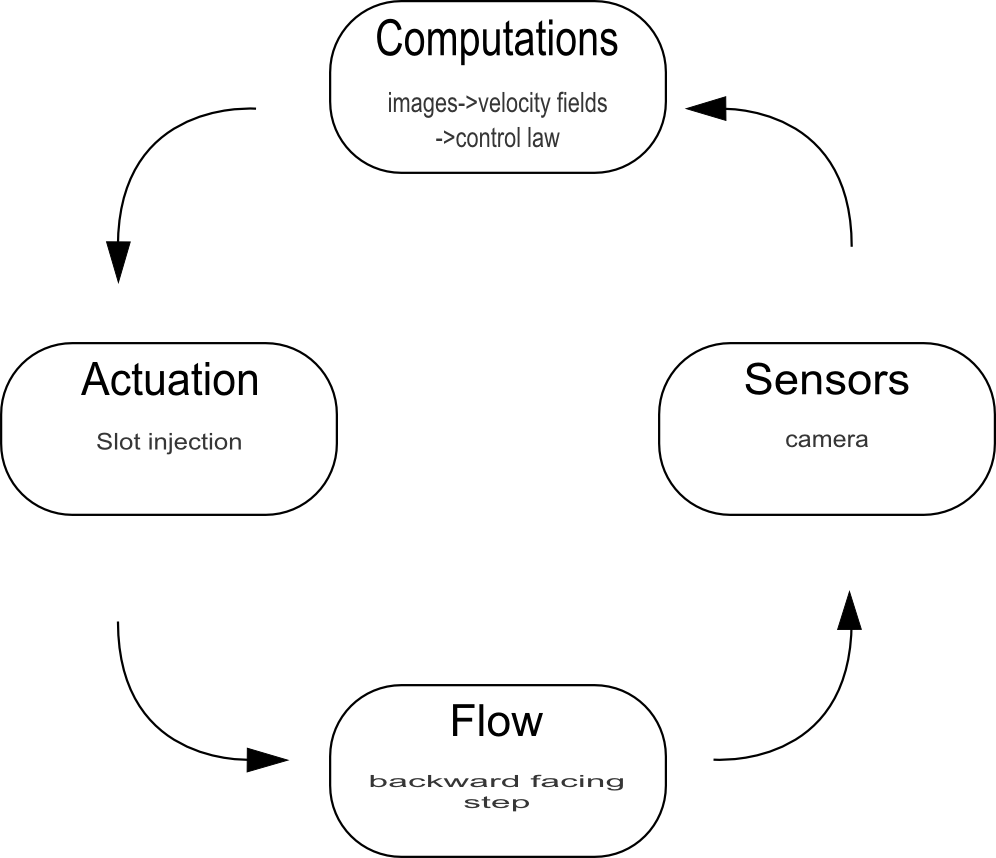}
\caption{Feedback scheme}
\label{fig:feedback}
\end{figure}

The flow is controlled using a spanwise, normal to the wall, slot jet, 0.1~cm long and 9~cm wide. The slot is located 3.5~cm upstream the step edge (figure~\ref{fig:dimensions}). Water coming from a pressurized tank enters a plenum and goes through a volume of glass beads designed to homogenize the incoming flow. Jet output is controlled by changing the tank pressure. The injection geometry was chosen to avoid 3D effects and keep the perturbation as bidimensional as possible. The control parameter, or manipulated variable, is the mean jet velocity $V_j$. Mean jet velocity varies from - 5 to 35~cm.s$^{-1}$. Since the jet supply tank is below the channel tank there is some suction when no pressure is supplied to the jet tank, allowing easy refilling of the tank. The dimensionless actuation amplitude is defined as the ratio of the mean jet velocity to the flow velocity $a_0 = \bar{V_j} / U_{\infty}$.
\\
The feedback loop is summarized in figure \ref{fig:feedback}.

In the following, open-loop experiments are carried out for all Reynolds numbers while closed-loop control experiments are carried out for one single Reynolds number, $Re_h=2900$. For this Reynolds number, shedding frequency does not exceed 2~Hz (\cite{Le1997}), thus any frequency higher than 20~Hz is sufficient to control the flow. The real-time velocimetry allows for feedback acquisition frequency up to 73~Hz for the SP case and 22.5~Hz for the HP case. The frequency is lower for the HP case because the images are bigger. It should be noted higher frequencies can be achieved for both cases but it was not needed for the current experiments.

\section{Characterization of the uncontrolled flow}
\subsection{Evolution of the recirculation bubble with $Re_h$}
\begin{figure}
\begin{center}
\begin{tabular}{c c}
\includegraphics[width=0.55\textwidth]{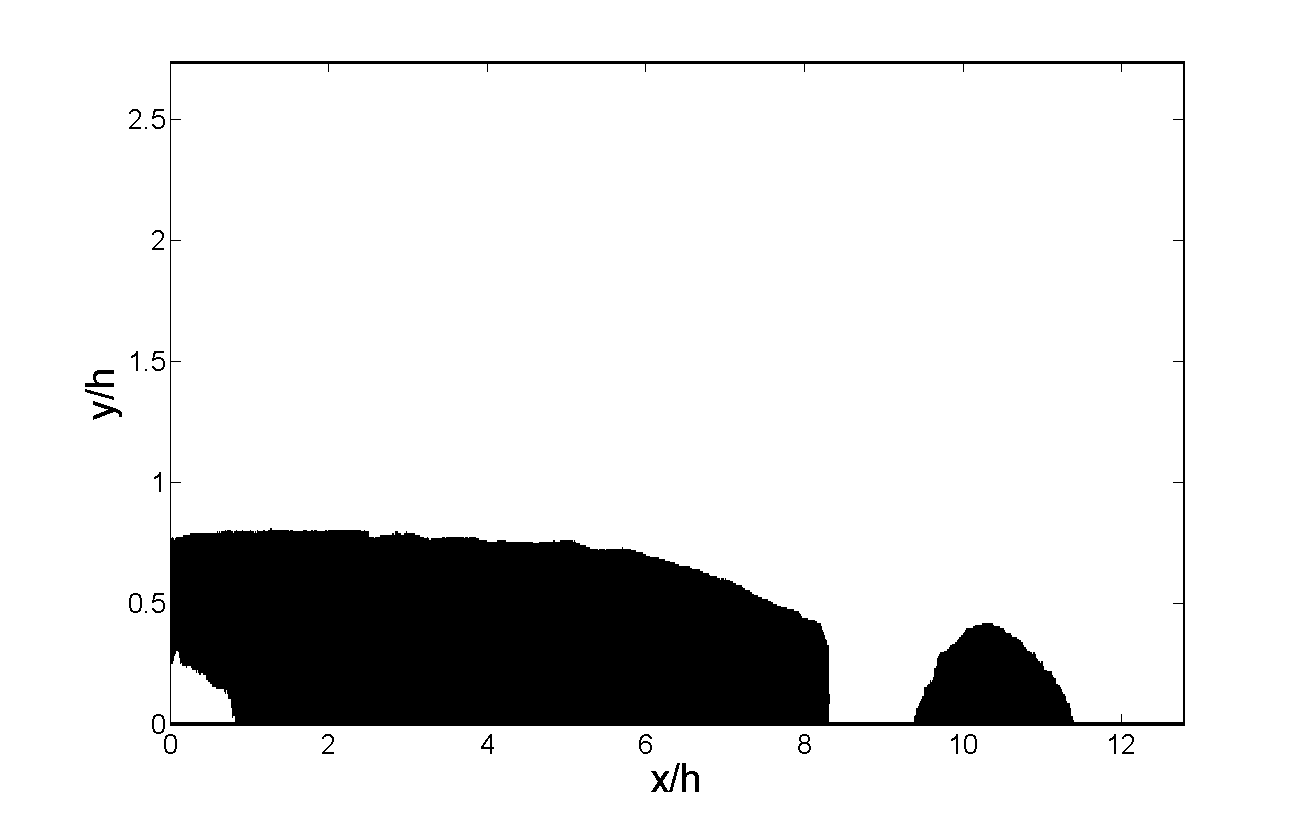} & \includegraphics[width=0.55\textwidth]{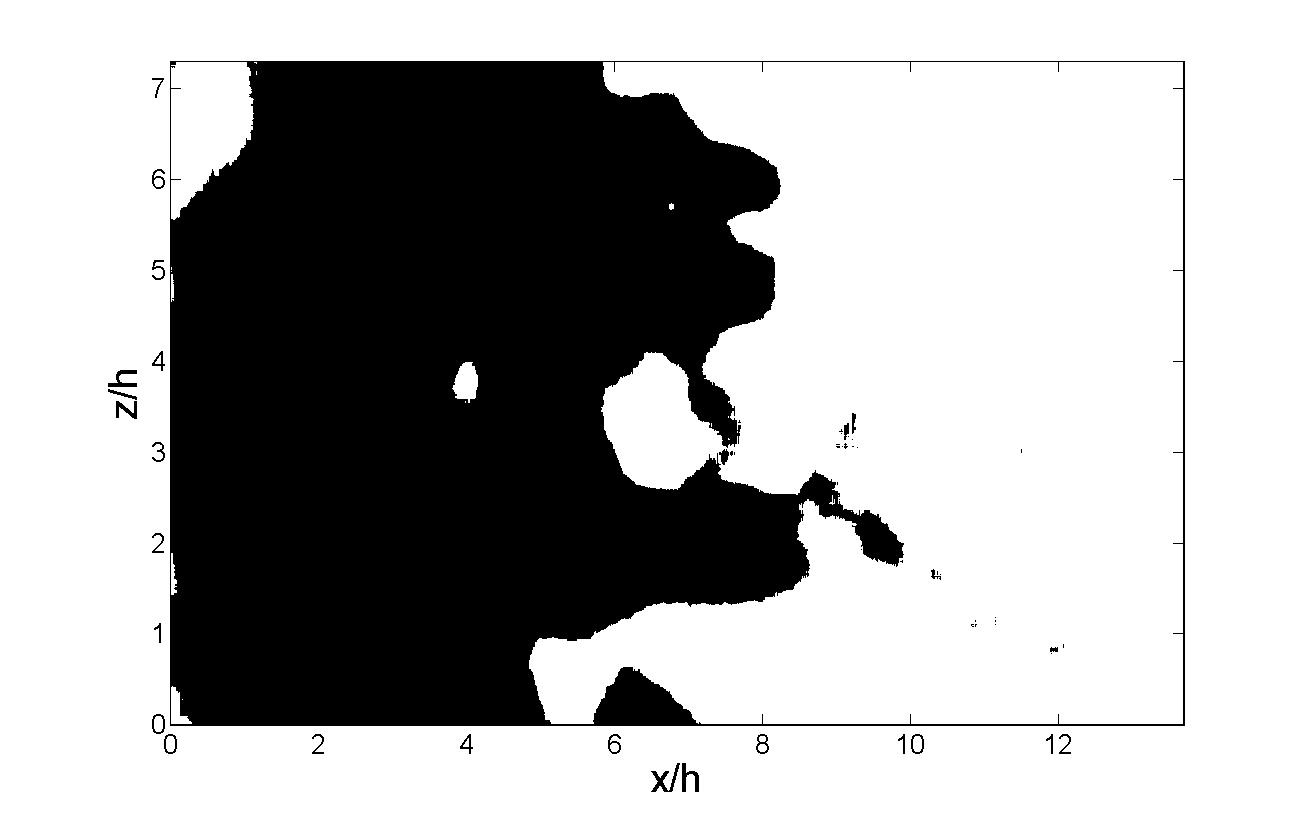}\\
a) & b)
\end{tabular}
\end{center}
\caption{a) Visualization of the instantaneous recirculation bubble surface (black region) for configuration SP (a) and HP (b).}
\label{fig:sbubble_side_above}
\end{figure}

The first step is to choose and define properly the quantity to be controlled. In the case of separated flows and more precisely backward-facing step flows, the length of the recirculation bubble $X_r$ is commonly used as input variable (\cite{King2007,Chun1996}). Because 2D velocity data are now available in the flow, the recirculation bubble can be characterized by its surface instead of its length.  The recirculation bubble surface can be considered to be the surface occupied by the region(s) of flow where longitudinal velocity is negative. Recirculation bubble surface $S_{bubble}$ is then defined in equation \ref{eq:surf}:

\begin{equation}
S_{bubble}(t)=\frac{1}{S}\int_{S} 1_{[v_x \le 0]} dS
\label{eq:surf}
\end{equation}

where $v_x(x,y)$ (respectively $v_x(x,z)$ for the HP case) is the streamwise velocity measured in the vertical $(x,y)$ (respectively $(x,z)$) plane. This definition presents several advantages. It is applicable regardless of camera position. It is simple, straightforward and can be implemented at low computational cost, thus computation does not slow down the feedback loop. Moreover no past data need be computed. Figure~\ref{fig:sbubble_side_above} shows instantaneous bubbles surfaces for both configurations. Black regions correspond to the separated flow. One can see that the regions are well-defined. The contours are irregular, with holes especially in the HP configuration. This is consistent with previous observation of instantaneous recirculation bubble (\cite{JLA2007}): the reattachment line is fully three-dimensional because of the destabilization of the transverse Kelvin-Helmholtz vortices shed in the shear layer. The scalar quantity used as an input for the closed-loop experiments can be $S_{bubble}(t)$, the spatial average of the regions in the instantaneous 2D velocity field where a backward flow is measured at time $t$.

It is also interesting to compute the time-averaged recirculation bubble length $<X_r>_t$ or surface $<S_{bubble}>_t$ as a function of the Reynolds number to compare with previous experimental or numerical studies. In the following, 1000 image pairs were taken at a sampling frequency  $F_s = 3.06$~Hz for 16 Reynolds numbers and for both configurations. 

\begin{figure}[H]
\centering
\begin{tabular}{c c}
\includegraphics[width=0.5\textwidth]{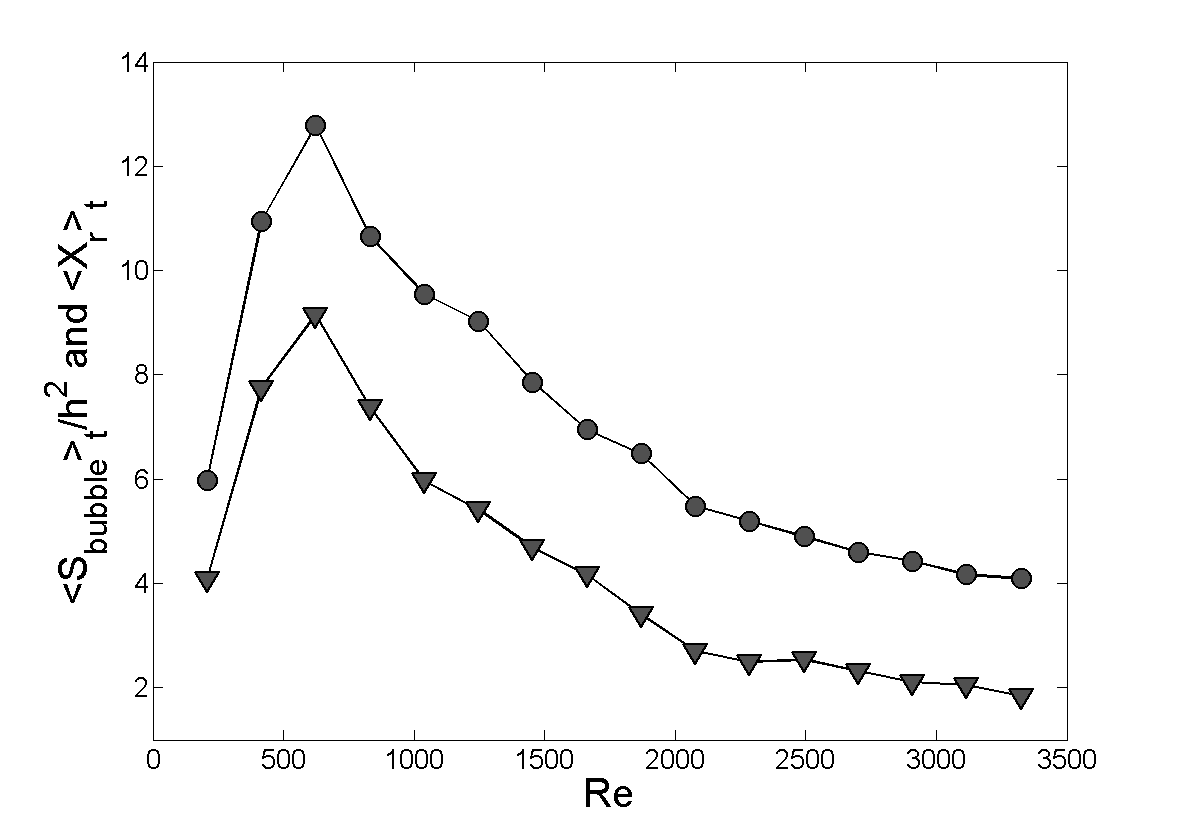} & \includegraphics[width=0.5\textwidth]{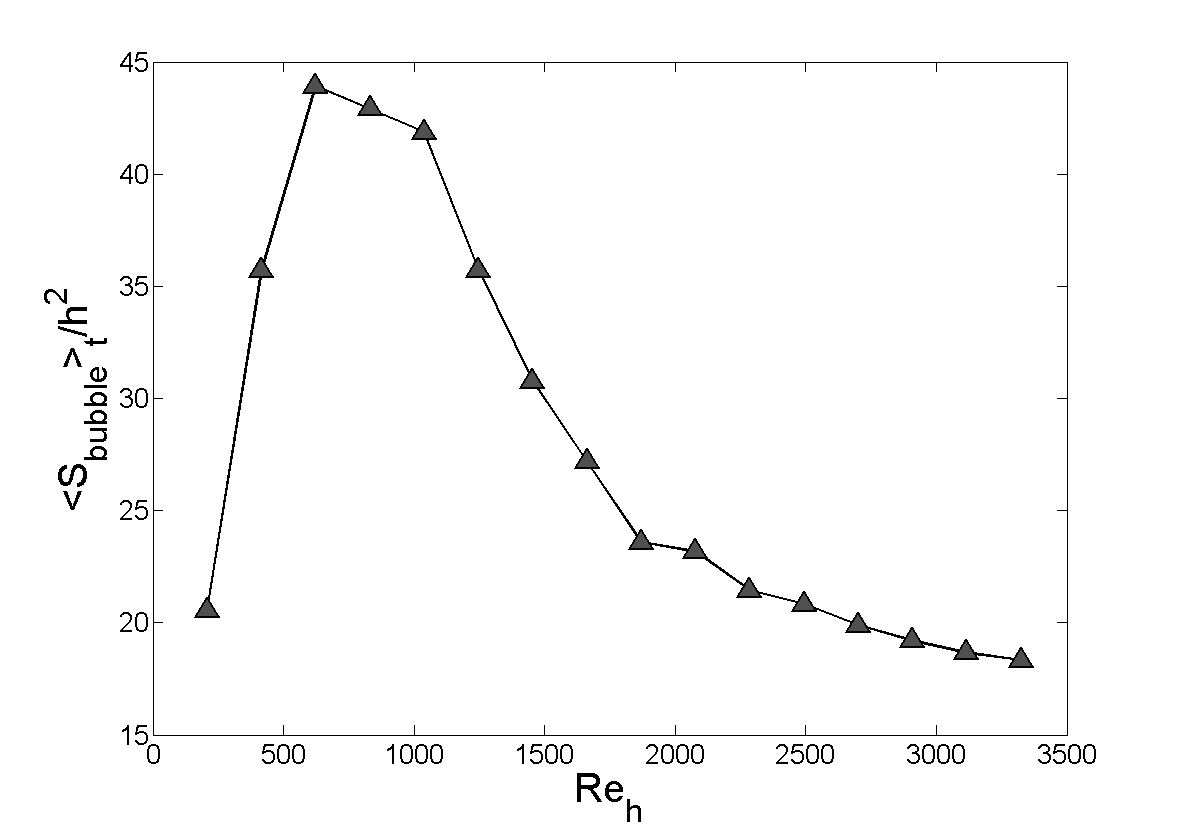}\\
a) & b)
\end{tabular}
\caption{a) Mean recirculation length (\FilledCircle) and mean recirculation surface (\FilledTriangleDown) as a function of $Re_h$ for the side configuration. b) Mean recirculation length (\FilledTriangleUp) as a function of $Re_h$ for the HP configuration.}
\label{fig:recirc_length}
\end{figure}

Figure~\ref{fig:recirc_length}a shows the evolution of mean recirculation bubble surface (normalized by $h^2$) for the SP configuration and mean recirculation bubble length $<X_r>_t$ (normalized by $h$) extracted from the mean longitudinal velocity field, choosing the second point at the wall where longitudinal velocity changes from negative to positive. Figure~\ref{fig:recirc_length}b shows the evolution of mean recirculation bubble surface (normalized by $h^2$) for the HP configuration. The normalized values are higher since the surface is effectively larger when observed from above, however the trend is similar. Thus observing the flow from above allows for the recovery of the recirculation bubble state. Moreover observing from above gives access to the span wise fluctuations of the bubble. It can be useful if one wishes to control the span wise reattachment, as was done by \cite{King2007} using a grid of 60 pressure sensors. 

Recirculation length evolves in a way consistent with previous observations (\cite{Armaly1983}). Furthermore, normalized recirculation bubble surface closely follows the evolution of bubble length, therefore making it a relevant parameter to characterize the flow state. These results also show how the evolution of the recirculation bubble can be followed whatever the plane the flow is observed. All this enables visual servoing to be used with any control scheme using the recirculation bubble surface as control variable.

\subsection{Evolution of the swirling intensity with $Re_h$}

A great advantage of visual servoing is that it also allows the experimentalist to compute previously inaccessible control variables, such as those proposed by \cite{Kunish1999}, i.e. variables involving quantities derived from the flow field such as velocity fluctuations, pressure fluctuations, and vorticity.\\

 We chose to compute the swirling strength criterion $\lambda_{ci}(s^{-1})$ which was first introduced by \cite{Chong1990} who analyzed the velocity gradient tensor and proposed that the vortex core be defined as a region where $\nabla u$ has complex conjugate eigenvalues.   It was later improved and used for the identification of vortices in three-dimensional flows by \cite{Zhou1999}. This criterion allows for an effective detection of vortices even in the presence of shear  (\cite{cambonie2013experimental}). The value $\frac{2 \pi}{\lambda_{ci}}$ at a given position is the time an element of fluid at this position would take to rotate around the nearest vortex core. 
 
\begin{figure}
\centering
\begin{tabular}{c c}
\includegraphics[height=0.3\textwidth]{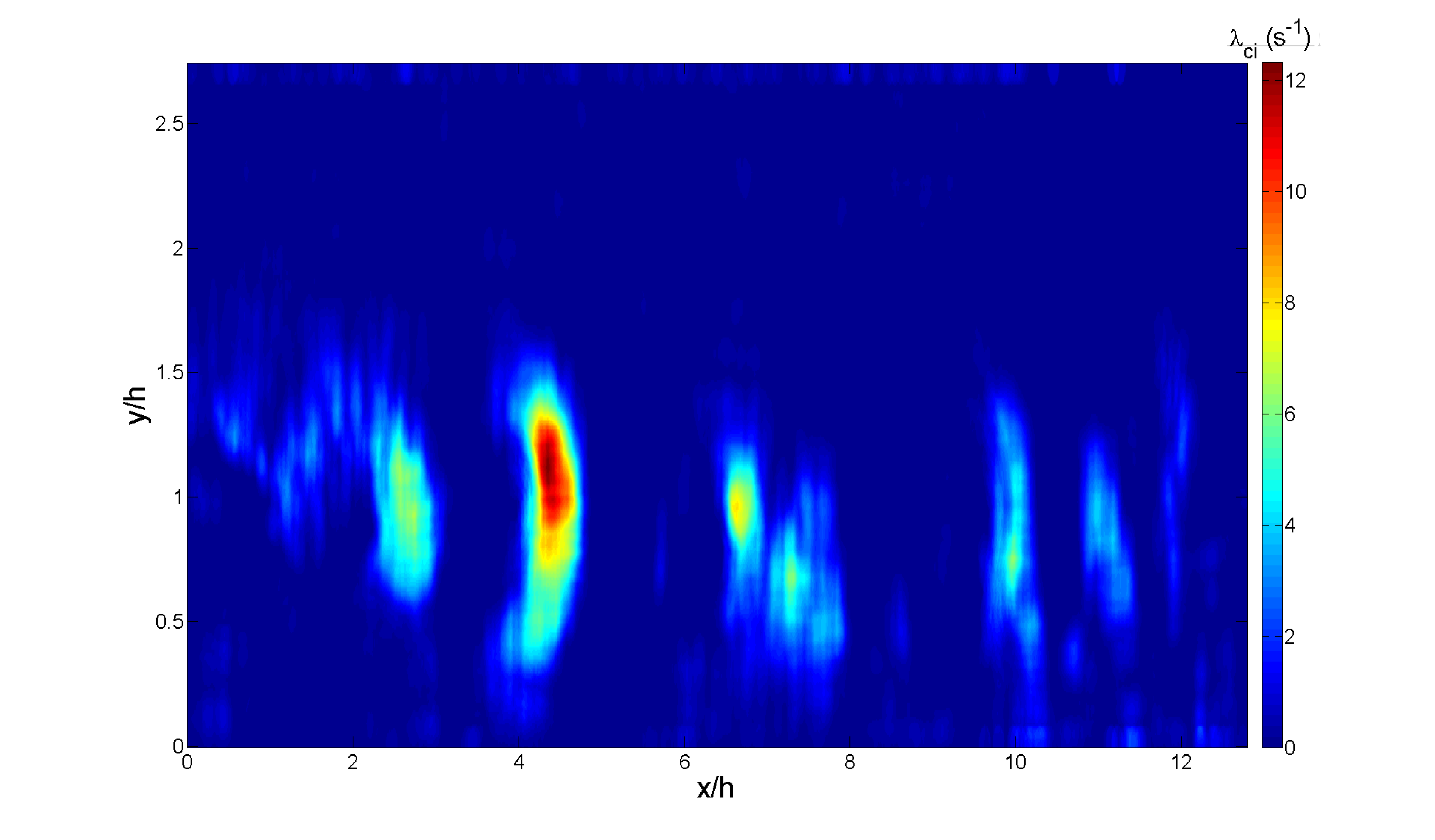} & \includegraphics[height=0.3\textwidth]{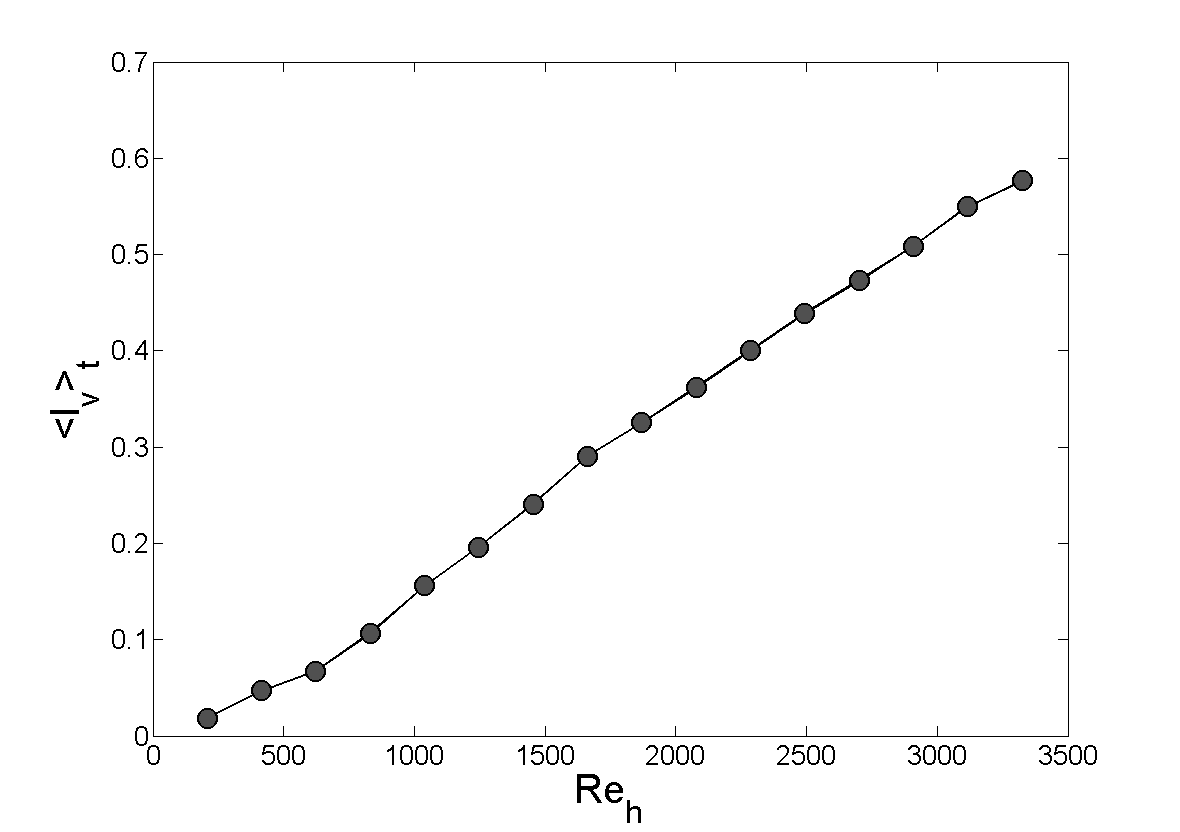} \\
a) & b)
\end{tabular}
\caption{a) Contour of instantaneous values of $\lambda_{ci}$ for $Re_h=2900$ . b) Time averaged values of $I_v(t)$ as a function of Reynolds number for the SP configuration. }
\label{fig:lci_Re}
\end{figure}

 Figure \ref{fig:lci_Re} shows an instantaneous map of $\lambda_{ci}$ for $Re_h=2900$. Regions of high swirling intensity are vortices created in the shear layer. It is then possible to spatially average $\lambda_{ci}$ to compute a scalar, hereafter noted $I_v(t)=\frac{1}{S}\int_S \lambda_{ci} (t) dS$, which effectively measures the combined intensity of the vortices present in the flow at a given time. Computation of $I_v(t)$ is implemented on the GPU to maintain high frequency sampling.  Figure~\ref{fig:lci_Re}b shows the evolution of spatially and time averaged swirling strength $<I_v>_t$ for the SP configuration. Only the SP configuration is used since it is best suited to detecting vortices  in the shear layer. The figure shows how the mean swirling strength increases with $Re_h$ until reaching a plateau.

\section{Open-loop experiments}
\begin{figure}[H]
\centering

\includegraphics[width=0.75\textwidth]{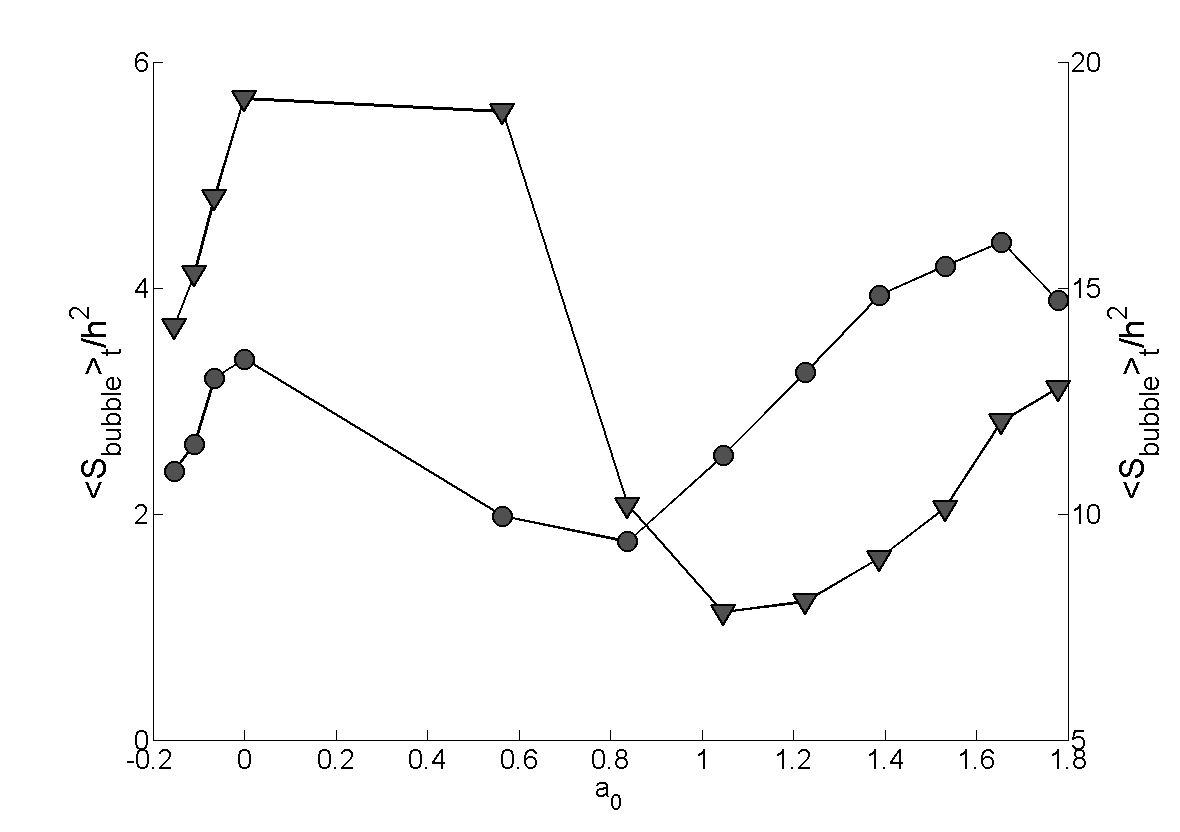}

\caption{ Mean recirculation length as a function of actuation amplitude for the SP configuration (\FilledCircle) and for the HP configuration (\FilledBigTriangleDown).}
\label{fig:recirc_open}
\end{figure}

Before turning to closed-loop experiments, it is important to characterize the open loop response of the system for both configurations, with $Re_h=2900$. It is a necessary step in order to choose a  proper closed-loop algorithm. Twelve actuations amplitudes were sampled for each configurations. Figure ~\ref{fig:recirc_open} shows the evolution of the mean recirculation length for both configurations as a function of actuation amplitude.

Both plots present the same qualitative characteristics: an increasing bubble size until jet output is null followed by a decrease to a minimum and a subsequent increase.  When actuation amplitude is negative (suction) the bubble becomes smaller, it grows in size as suction diminishes. Once amplitude becomes positive the flow remains unperturbed until the jet is strong enough the affect the bubble. Once this happens bubble surface quickly diminishes. In both cases, the recirculation bubble is minimum when $a_0 \approx 1$, i.e. when the jet velocity is close to the freestream velocity. However when the amplitude of the actuation becomes strong enough, it creates a fluid wall, essentially protecting the recirculation region from the incoming flow, causing bubble surface to increase once again. Since the relation between bubble size and actuation amplitude presents a global minimum ($-60\%$ in both cases), a gradient descent type control algorithm is advisable.

\begin{figure}[H]
\begin{center}
\includegraphics[width=0.7\textwidth]{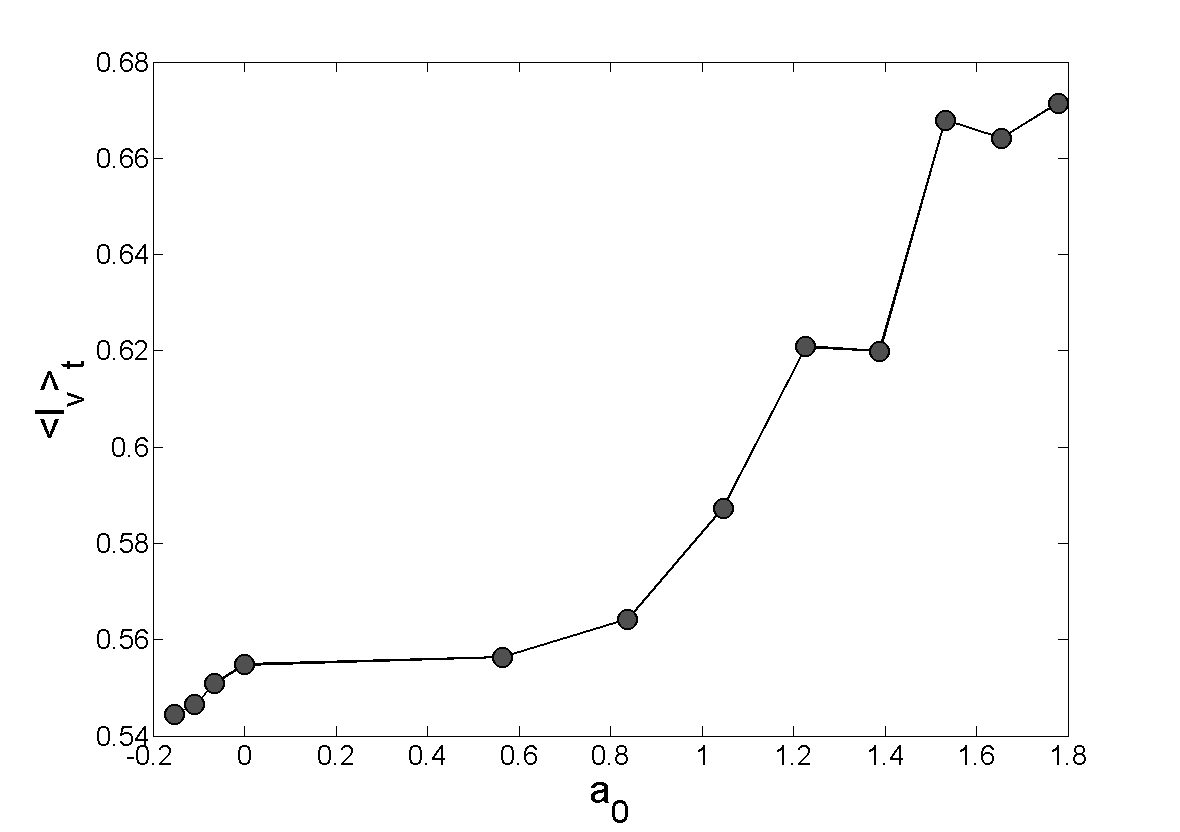}
\caption{Evolution of $<I_v>_t$ with actuation amplitude for $Re_h=2900$.}
\end{center}
\label{fig:open_lci}
\end{figure}

Figure~\ref{fig:open_lci} shows the spatially and time averaged values of swirling strength $<I_v>_t$ as a function of jet intensity. It increases with actuation amplitude. At first, actuation reinforces the vortices present in the shear layer. Once the jet is strong enough to create vortices outside of the shear layer the strength of these vortices is added to the vortices created in the shear layer resulting in an overall higher swirling intensity $I_v$.

\section{Closed-loop experiments}
\subsection{Gradient-descent algorithm}
As mentioned previously, the evolution of the manipulated variable $S_{bubble}$ as a function of actuation amplitude exhibits a clear minimum. It is thus well-suited to a gradient-descent control such as slope-seeking (\cite{Krstic2000,Beaudoin2006}). However, it is not possible with the present experimental setup to add a periodic excitation. A simpler, albeit less robust, implementation of gradient-descent will be used in order to demonstrate the feasibility and advantages of visual servoing when controlling a separated flow. The algorithm used is a basic gradient descent algorithm. Actuation is changed iteratively in opposition to the controlled variable's slope.

\begin{figure}
\centering
\begin{tabular}{c c }
\includegraphics[width=0.5\textwidth]{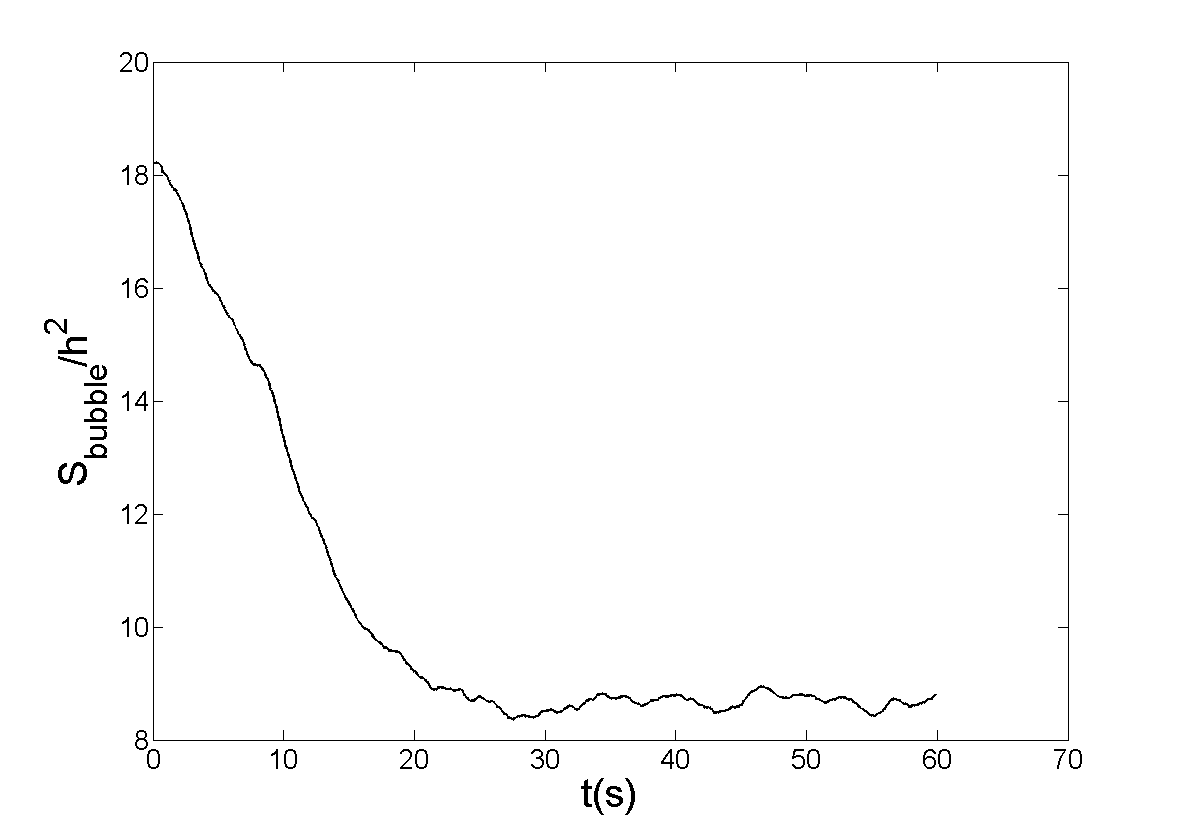}  & \includegraphics[width=0.5\textwidth]{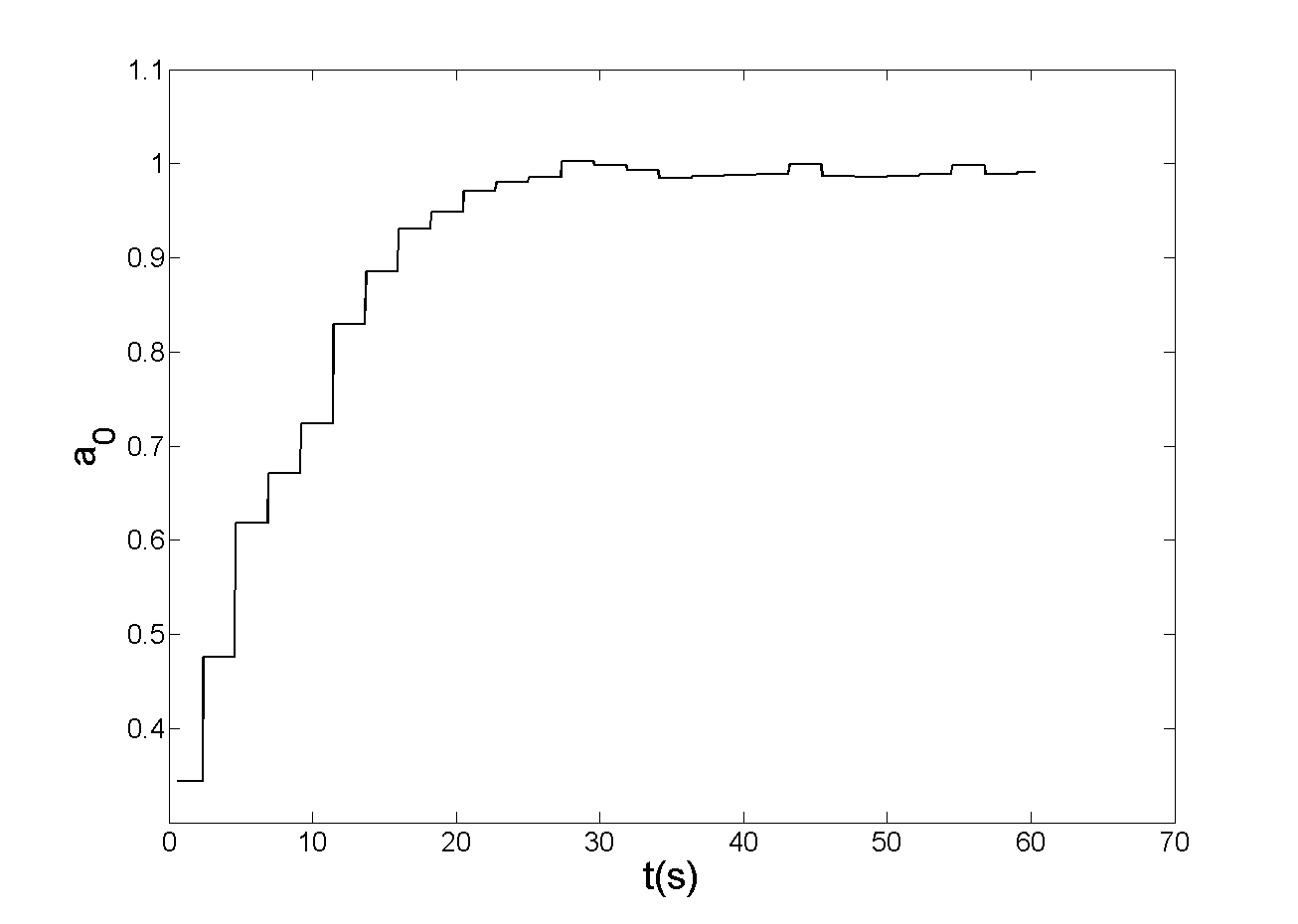}\\
a) & b)
\end{tabular}
\caption{a) Closed loop bubble surface during gradient descent for the HP configuration at $Re_h=2900$. b) Corresponding evolution in actuation amplitude.}
\label{fig:Grad_control}
\end{figure}

Figure~\ref{fig:Grad_control}a shows the evolution of $S_{bubble}$ as a function of time during minimum seeking.  Figure~\ref{fig:Grad_control}b shows the corresponding evolution of actuation amplitude. One can see that when actuation starts, $S_{bubble}$ decreases regularly until reaching a minimum (giving $-60\%$ reduction) after 20 seconds. The system remains in this state as long as actuation is applied. Actuation amplitude increases regularly until reaching a plateau at $a_0=0.95$ which corresponds to the optimal amplitude leading to the minimum recirculation bubble obtained in the open-loop experiment.

This clearly demonstrates how a visual feedback system can be used to implement the same control methods as those using recirculation bubble length as control variable. Convergence speed is 20 seconds, this could be improved by improving the control algorithm.

\subsection{PID control}
\begin{figure}[H]
\centering
\includegraphics[width=0.5\linewidth]{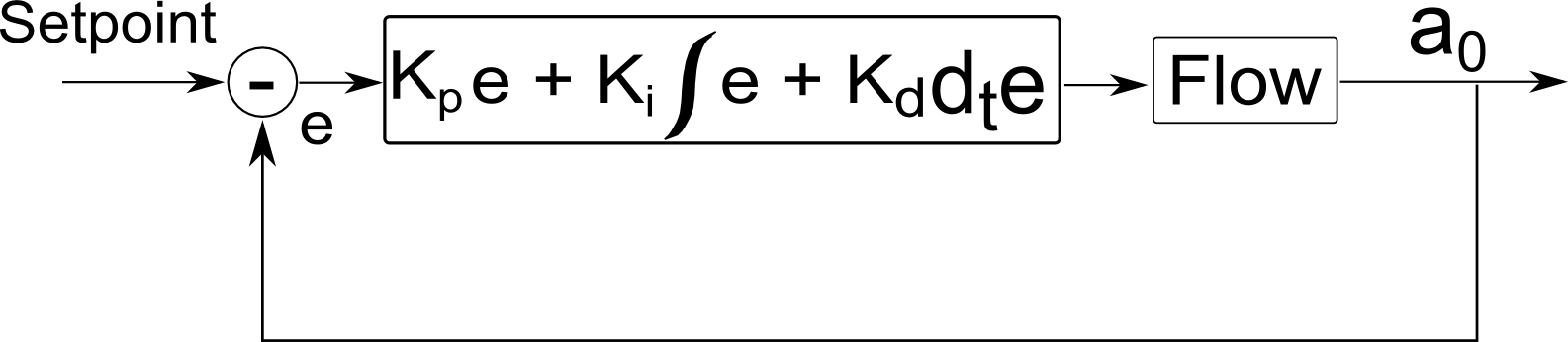}
\caption{PID algorithm in block form.}
\label{fig:PID_block}
\end{figure}

Because $<I_v>_t$ is a monotonous function of the actuation amplitude, it is well suited to control via a PID algorithm. The PID (Proportional-Integral-Derivative) controller is fundamental in control theory and is very useful when no model of the system is available. For informations on PID control, see \cite{Hagglund2001}. Essentially, an arbitrary setpoint command is given to the controller which then computes the appropriate actuation required to bring the system to a state giving the desired setpoint value. Following the methodology described in \cite{Atherton1993}, a automatically tuned PID controller was implemented. The system is brought to a state of controlled oscillation and optimal PID parameters are computed.  The PID algorithm is detailed in figure \ref{fig:PID_block}.

\begin{figure}
\centering
\begin{tabular}{c c }
\includegraphics[width=0.5\textwidth]{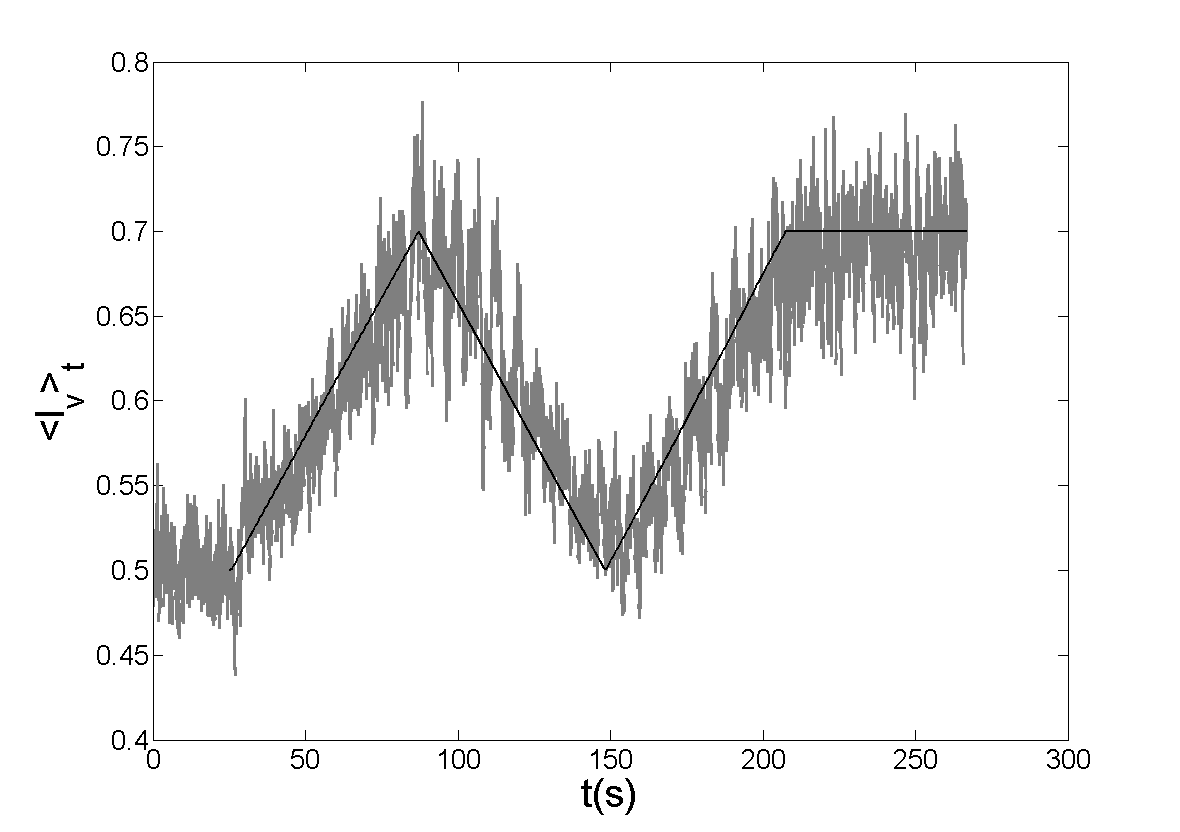} & \includegraphics[width=0.5\textwidth]{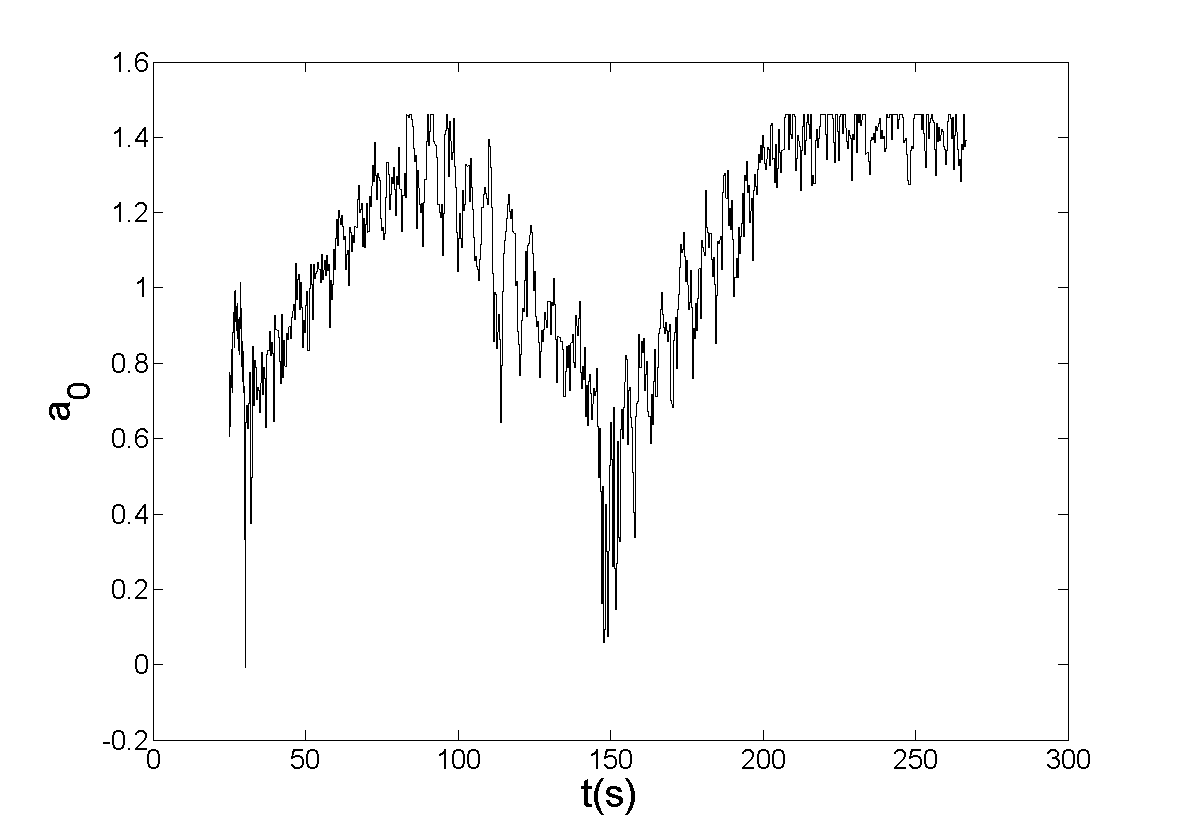}\\
a) & b)

\end{tabular}
\caption{a) Closed loop values of $I_v(t)$ during PID control for configuration SP and $Re_h=2900$. Control variable in grey, setpoint in black. b) Corresponding evolution in actuation amplitude. }
\label{fig:PID_control}
\end{figure}

Figure~\ref{fig:PID_control}a shows the evolution of the control variable during PID control and figure~\ref{fig:PID_control}b the corresponding changes in actuation amplitude. These figures show how mean swirling strength can be drastically changed in less than 60 seconds.

\section{Conclusion}
An experimental study of control by visual feedback on the detached flow downstream a backward facing-step has been carried out in an hydrodynamic channel. High frequency, low latency computations of the velocity field behind the step were used to define and compute two novel control variables: recirculation bubble surface and mean swirling strength. The evolution of these variables as a function of the Reynolds number and as a function of actuation amplitude for a given Reynolds number have been studied.\\
The surface of the recirculation bubble is shown to behave in much the same way as its length. Hence visual informations can be used to control the BFS flow using standard feedback schemes. Since the open loop evolution of the the recirculation bubble surface presents a minimum when the actuation amplitude is varied, a gradient descent algorithm has been chosen and successfully implemented in the feedback loop.\\
Thanks to the real-time velocity measurements, new control variables can be defined. The spatially averaged swirling strength allows for the estimation of the intensity of the vortices created in the separated flow. This is the first time such a variable is computed in real-time from online flow velocity data, demonstrating the new avenues opened for control  by visual servoing. The open loop response shows a relatively smooth evolution of the mean swirling strength as a function of the actuation amplitude, well fitted for a PID controller.  A closed loop implementation demonstrates how swirling strength in a detached flow can be dynamically controlled  through visual feedback.

\newpage
\bibliographystyle{unsrt}	
\bibliography{Bibliography}

\begin{thebibliography}{10}

\bibitem{King2007}
L.~Henning and R.~King.
\newblock Robust multivariable closed-loop control of a turbulent
  backward-facing step flow.
\newblock {\em Journal of Aircraft}, 44, 2007.

\bibitem{CHC}
R.L. Simpson.
\newblock Aspect of turbulent boundary layer separation.
\newblock {\em Progress in Aerospace Sciences}, {32}:457--521, 1996.

\bibitem{Hucho2005}
W.D. Hucho.
\newblock {\em Aerodynamic of Road Vehicles}.
\newblock Vieweg, 2005.

\bibitem{Weisenstein2000}
C.~O Paschereit, E.~Gutmark, and W.~Weisenstein.
\newblock Excitation of thermoacoustic instabilities by the interaction of
  acoustics and unstable swirling flow.
\newblock {\em AIAA Journal}, 38:1025--1034, 2000.

\bibitem{Fernholz1990}
H.~Fiedler and H.H. Fernholz.
\newblock On the management and control of turbulent shear flows.
\newblock {\em Prog. Aerospace Sci.}, 72:305--387, 1990.

\bibitem{Chun1996}
K.~B. Chun and H.~J. Sung.
\newblock Control of turbulent separated flow over a backward-facing step by
  local forcing.
\newblock {\em Experiments in Fluids}, 21:417--426, 1996.

\bibitem{Mazur2007}
V.~Uruba, P.~Jonas, and O.~Mazur.
\newblock Control of a channel-flow behind a backward-facing step by
  suction/blowing.
\newblock {\em Heat and Fluid Flow}, 28:665--672, 2007.

\bibitem{Beaudoin2006}
J.-F. Beaudoin, O.~Cadot, J.-L. Aider, and J.~E. Wesfreid.
\newblock Drag reduction of a bluff body using adaptive control methods.
\newblock {\em Physics. Fluids}, 18:1, 2006.

\bibitem{Pastoor2008}
Mark Pastoor, Lars Henning, Bernd~R. Noack, Rudibert King, and Gilead Tadmor.
\newblock Feedback shear layer control for bluff body drag reduction.
\newblock {\em Journal of Fluid Mechanics}, 608:161--196, 2008.

\bibitem{Armaly1983}
B.~F. Armaly, F.~Durst, J.~C.~F. Pereira, and B.~Schonung.
\newblock Experimental and theoretical investigation of backward-facing step
  flow.
\newblock {\em Journal of Fluid Mechanics}, 127:473--496, 1983.

\bibitem{Le1997}
L.~Hung, M.~Parviz, and K.~John.
\newblock Direct numerical simulation of turbulent flow over a backward-facing
  step.
\newblock {\em Journal of Fluid Mechanics}, 330:349--374, 1997.

\bibitem{JLA2004}
J-F. Beaudoin, O.~Cadot, J-L. Aider, and J.E. Wesfreid.
\newblock Three-dimensional stationary flow over a backwards-facing step.
\newblock {\em European Journal of Mechanics}, 38:147--155, 2004.

\bibitem{JLA2007}
J-L. Aider, A.~Danet, and M.~Lesieur.
\newblock Large-eddy simulation applied to study the influence of upstream
  conditions on the time-dependant and averaged characteristics of a
  backward-facing step flow.
\newblock {\em Journal of Turbulence}, 8, 2007.

\bibitem{Sipp2010}
D.~Sipp, A.~Barbagallo, and P.~Schmid.
\newblock Closed-loop control of an unstable open cavity.
\newblock {\em Journal of Fluid Mechanics}, 641:1--50, 2010.

\bibitem{Collewet2011}
R.T. Fomena and C.~Collewet.
\newblock Fluid flow control : A vision-based approach.
\newblock {\em International Journal of Flow Control}, pages 133--169, 2011.

\bibitem{Willert2010}
M.~Gharib C.~Willert, M.~Munson.
\newblock Real-time particle image velocimetry for closed-loop flow control
  applications.
\newblock 15th international Symposium on Applications of laser techniques to
  fluid mechanics, 05-08 Jul. 2010. Lisbon Portugal, 2010.

\bibitem{Roberts2012}
J.~Roberts.
\newblock {\em Control of underactuated fluid-body systems with real-time image
  velocimetry}.
\newblock Phd, MIT, 2012.

\bibitem{Gautier2013OF}
N.~Gautier and J-L. Aider.
\newblock Real time, high frequency planar flow velocity measurements.
\newblock {\em Submitted Open Journal of Fluid Dynamics, avalaible at arXiv
  under Real-time planar flow velocity measurements using an optical flow
  algorithm implemented on GPU}, 2013.

\bibitem{Lucas1984}
Bruce~D. Lucas.
\newblock {\em Generalized image matching by the method of differences}.
\newblock Phd, Carnegie Mellon University, 1984.

\bibitem{Plyer2011}
F.~Champagnat, A.~Plyer, G.~Le~Besnerais, B.~Leclaire, S.~Davoust, and
  Y.~Le~Sant.
\newblock Fast and accurate piv computation using highly parallel iterative
  correlation maximization.
\newblock {\em Experiments in Fluids}, 50:1169--1182, 2011.

\bibitem{Kunish1999}
H.~Choi, M.~Hinze, and K.~Kunisch.
\newblock Instantaneous control of backward-facing step flows.
\newblock {\em Applied Numerical Mathematics}, 31:133--158, 1999.

\bibitem{Chong1990}
M.S. Chong, A.E. Perry, and B.J. Cantwell.
\newblock {A general classification of 3-dimensional flow fields}.
\newblock {\em {Physics of Fluids}}, {2}:{765--777}, {1990}.

\bibitem{Zhou1999}
J.~Zhou, R.J. Adrian, S.~Balachandar, and T.M. Kendall.
\newblock Mechanisms for generating coherent packets of hairpin vortices.
\newblock {\em J Fluid Mech}, 387:535--396, 1999.

\bibitem{cambonie2013experimental}
T~Cambonie, N~Gautier, and J-L Aider.
\newblock Experimental study of counter-rotating vortex pair trajectories
  induced by a round jet in cross-flow at low velocity ratios.
\newblock {\em Experiments in Fluids}, 54(3):1--13, 2013.

\bibitem{Krstic2000}
Miroslav Krstic.
\newblock Performance improvement and limitations in extremum seeking control.
\newblock {\em Systems \& Control Letters}, 39:313--326, 2000.

\bibitem{Hagglund2001}
KJ. Astrom and T.~Hagglund.
\newblock The future of pid control.
\newblock {\em Control Engineering Practice}, 9:1163--1175, 2001.

\bibitem{Atherton1993}
M.~Zhuang and D.P. Atherton.
\newblock Automatic tuning of optimum pid controllers.
\newblock {\em IEEE Proc Control Theory and Appl}, 140, 1993.

\end{thebibliography}

\end{document}